\documentclass[12pt]{article}
\usepackage{amsmath}

\begin{document}

%&&&& definitions &&&&&&
\newcommand{\dx}{\mathbf{d}x\,}
\newcommand{\dk}{\mathbf{d}k\,}
\newcommand{\Op}{\mathcal{O}}
\newcommand{\hp}{\hat{\phi}}
\newcommand{\cp}{\check{\phi}}
\newcommand{\e}[1]{\mathrm{e}^{#1}}

%%%%% end of definitions %%%%%

\begin{titlepage}

\begin{flushright}
hep-th/0201100\\
DSF--2--2002
\end{flushright}\vspace{1.5cm}

\begin{center}
\LARGE \textbf{An improved correspondence formula for AdS/CFT with
multi-trace operators}
\end{center}\vspace{1cm}

\begin{center}
\large Wolfgang M\"{u}ck
\end{center}

\begin{center}
\emph{Dipartimento di Scienze Fisiche, Universit\`{a} di Napoli
``Federico II'' and INFN, Sezione di Napoli,
Via Cintia, 80126 Napoli, Italy}\\
\texttt{mueck@na.infn.it}
\end{center}\vspace{1.5cm}

\begin{abstract}
An improved correspondence formula is proposed for the calculation of
correlation functions of a conformal field theory perturbed by
multi-trace operators from the analysis of the dynamics of the dual
field theory in Anti-de Sitter space. The formula reduces to the usual
AdS/CFT correspondence formula in the case of single-trace
perturbations.
\end{abstract}

\end{titlepage}

In a recent paper \cite{Witten02a} Witten proposed an improved boundary
condition to be used in the AdS/CFT correspondence when the boundary
conformal field theory is perturbed by multi-trace operators,
\begin{equation}
\label{action}
  I_{\text{QFT}}[\Phi] = I_{\text{CFT}}[\Phi] + \int \dx W[\Op(\Phi)]~,
\end{equation}
where $\Phi$ denotes a set of fundamental fields, $\Op$ is a scalar
primary operator, $W[\Op]$ is an arbitrary function of $\Op$, and
$\dx$ stands for the covariant 
volume integral measure. In the case of $W[\Op] = \frac12 f \Op^2$, and
$\Op$ of conformal dimension $\Delta= d/2$, Witten demonstrated that the
proposed boundary condition yields the expected renormalization formula
for the coupling $f$. One would expect that the AdS/CFT correspondence
also yields the two-point function in the presence of a finite
coupling, $f$, which is related to the Green function in the absence
of $f$ by Dyson's formula. However, it is not difficult
to realize that the usual AdS/CFT correspondence formula
\cite{Gubser98-1,Witten98-1},
$\exp(-I_{\text{AdS}}) =\langle \exp(-\int \dx \alpha \Op) \rangle$,
where $I_{\text{AdS}}$ is the regularized and renormalized bulk
on-shell action, and $\alpha$ is a generating current, does not
reproduce this result.

Double-trace deformations, modified boundary conditions and their
relations to non-local string theory have been studied recently in
\cite{Berkooz02a}. The discussion in that paper differs from Witten's
and the present one in that the change of boundary conditions appears in the
un-renormalized bulk action, whereas in our approach holographic
renormalization is carried out as usual.\footnote{For a recent
systematic description of holographic renormalization see
\cite{Bianchi01b}.} 

The purpose of the present letter is to present an improved
correspondence formula, which gives expected boundary field theory
correlators for multi-trace perturbations. The new method will
naturally apply to both regular and irregular boundary conditions of
the bulk fields \cite{Breitenlohner82}, which correspond to field
theory operators of dimensions $\Delta=d/2+\lambda$ and
$\Delta=d/2-\lambda$, respectively \cite{Klebanov99-1}. ($\lambda$ is
positive and in the second case satisfies the unitarity bound
$\lambda<1$.) 

We shall for simplicity consider a single scalar field in AdS bulk
space and first explain the general method. Later, we will test the
formula for a free bulk field and 
$W=\int \dx (\beta \Op +\frac12 f \Op^2)$, where $\beta$ is a finite
soucre, and $f$ is a coupling constant. 

First, to clarify our notation, we shall consider an AdS bulk of
dimension $d+1$ with a metric  
\begin{equation}
\label{metric}
  ds^2 = r^{-2} (dr^2 + d x^2)~, 
\end{equation}
where $dx^2$ denotes the Euclidean metric in $d$ dimensions. The
asymptotic boundary (horizon) is located at $r=0$. For small $r$, a
scalar field behaves asymptotically as 
\begin{equation}
\label{asympt}
  \phi(r,x) = r^{d/2-\lambda} \frac1{2\lambda} \hp(x) 
  + r^{d/2+\lambda} \cp(x) +\cdots~,
\end{equation}
where $\lambda$ is related to the mass of the bulk field by $\lambda
= \sqrt{d^2/4+m^2}$.\footnote{For $\lambda=0$, the formula
\eqref{asympt} degenerates to $\phi(r,x) = r^{d/2} ( \hp \ln r +\cp )
+\cdots$.}  For later convenience, we have included the normalization factor
$1/(2\lambda)$ in the leading term. The two independent series solutions are
determined by $\hp$ and $\cp$, which are called the regular and
irregular boundary data \cite{Breitenlohner82}, respectively, and the
ellipses denote all subsequent terms of the two series. Regularity of the bulk
solution as well as bulk interactions determine the relation between
$\hp$ and $\cp$ uniquely. 

The main outcome of the AdS bulk analysis \cite{Gubser98-1,Witten98-1}
is the regularized and renormalized bulk on-shell action as a
functional of the regular boundary data, $I[\hp]$. Moreover, the two
boundary data $\hp$ and $\cp$ are canonically conjugate and satisfy
\cite{Klebanov99-1}
\begin{equation}
\label{canon}
  \cp = - \frac{\delta I[\hp]}{\delta \hp} \quad \text{and} \quad 
  \hp = \frac{\delta J[\cp]}{\delta \cp}~,
\end{equation}
where 
\begin{equation}
\label{Jdef}
  J[\cp] = I - \int \dx \hp \frac{\delta I[\hp]}{\delta \hp}
\end{equation}
is the Legendre transform of $I$. It has been proven in
\cite{Mueck99-4} that the formulae \eqref{canon} and \eqref{Jdef} hold
for interacting bulk fields to any order in perturbation theory.
Armed with these results, we can proceed to construct a generating
functional of the boundary field theory. 

In order to calculate correlation functions in the quantum field
theory with an action $I_{\text{QFT}}$, we need to add to
$I_{\text{QFT}}$ a source term, $\int \dx \alpha \Op$, which 
can be viewed as a single-trace perturbation of the theory. 
In the usual AdS/CFT calculation [$W=0$ in eqn.\ \eqref{action}], one
identifies one of the boundary data $\hp$ or $\cp$ with the source
$\alpha$. As a result [c.f.\ eqn.\ \eqref{canon}] the other one
corresponds to the boundary field theory operator, $\Op$. It turns out
that this second property is the fundamental one and can be used when
multi-trace perturbations are switched on. 
Let us first assume that $\Op$ corresponds to $\cp$. Then, the
perturbation of the conformal field theory, $\int \dx (W[\Op] + \alpha
\Op)$, translates into a functional of $\cp$, and we shall add it to
$J[\cp]$ in order to construct a generating functional,
\begin{equation}
\label{S} 
  S = J[\cp] + \int \dx (W[\cp] + \alpha\cp )~. 
\end{equation}
The choice of $J[\cp]$ is natural, since it is a functional of the
same data as the perturbation. The relation between the field $\cp$
and the source $\alpha$ is determined by demanding
\begin{equation}
\label{bc}
  \frac{\delta S}{\delta \cp} = \frac{\delta J}{\delta \cp} +
  \frac{dW}{d\cp} + \alpha = 0~.
\end{equation}
By virtue of eqn.\ \eqref{canon}, this is identical to Witten's
improved boundary condition \cite{Witten02a} (up to a sign that
depends on the convention for the Green function), and it reduces to the
regular boundary condition for $W=0$. Furthermore, we can re-write $S$
as 
\[ S = S - \int \dx \cp \frac{\delta S}{\delta \cp} = I + \int \dx
\left( W - \cp \frac{dW}{d\cp} \right)~, \]
so that $S$ is identical to the bulk on-shell action, $I$,  for linear $W$
(single-trace perturbations).   

Solving eqn.\ \eqref{bc} for $\cp$ and substituting it into eqn.\
\eqref{S} yields a functional $S[\alpha]$, which we identify as the
generating functional of the boundary field
theory. More precisely, we identify the bulk analysis with the
execution of the field theory functional integral, \emph{i.e.}, 
\begin{equation}
\label{corr1}
  \exp(-S[\alpha]) = \int \mathcal{D} \Phi\, 
  \exp \left\{ - \left[ I_{\text{CFT}}[\Phi] + 
  \int \dx (W[\Op] + \alpha\Op )\right] \right\}~,
\end{equation}
so that 
\begin{equation}
\label{corr}
  \exp[-(S[\alpha]-S[0])] = \left\langle \exp\left(-\int \dx \alpha
  \Op \right) \right\rangle_W~.
\end{equation}
Here, the expectation value is in the theory including the
perturbation $W$. Notice that eqn.\ \eqref{corr} ensures that $\langle
1 \rangle_W =1$. 

It is now obvious how to obtain the irregular boundary
condition. Identifying the boundary field theory operator $\Op$ with
$\hp$, we define $S$ by 
\begin{equation}
\label{S2} 
  S = I[\hp] + \int \dx (W[\hp] + \alpha\hp )~.
\end{equation}
The rest of the analysis goes through as before. In particular,
for $W=0$ we have $\cp=\alpha$ and $S[\alpha]= J[\cp]$, which is the
correct result for irregular boundary conditions
\cite{Klebanov99-1,Mueck99-4}.     

Let us check the improved correspondence formula by considering a free
bulk scalar of mass $m$ and a perturbation
\begin{equation}
\label{perturb}
  W[\Op]  =  \beta(x) \Op + \frac{f}2 \Op^2~.
\end{equation}
The AdS bulk analysis yields
\cite{Klebanov99-1,Mueck99-4}  
\begin{equation}
\label{exI} 
  I = -\frac12 \int \dx \hp(x) \cp(x)~,
\end{equation}
and regularity of the bulk solution dictates that $\cp$ and $\hp$ are
related by (in momentum space) 
\begin{equation}
\label{excphp} 
  \cp(k) = - \frac{\Gamma(1-\lambda)}{\Gamma(1+\lambda)}
  \left(\frac{|k|}2 \right)^{2\lambda} \hp(k) = G(k) \hp(k)~.
\end{equation}
Here, $G(k)$ denotes the conformal Green function in momentum space for
primary fields of dimension $\Delta=d/2+\lambda$.  
Hence, we find from eqns.\ \eqref{Jdef} and \eqref{excphp},  
\begin{equation}
\label{exJ}
  J[\cp] = \frac12 \int \dk \cp(k) \frac1{G(k)} \cp(-k)~.
\end{equation}
Thus, eqn.\ \eqref{S} yields 
\begin{equation}
\label{exS}
  S =  \int \dk \left\{ \cp(k) \frac12 \left[ \frac1{G(k)} +f \right] \cp(-k) +
  [\alpha(k)+\beta(k)]  \cp(-k) \right\}~,
\end{equation}
and the boundary condition is 
\begin{equation}
\label{exbc}
   \cp(k) = - [\alpha(k) +\beta(k)] \frac{G(k)}{1 + f G(k)}~.
\end{equation}
Inserting $\cp$ into $S$ yields 
\begin{equation}
\label{exS2}
  S[\alpha] -S[0] = -\frac12 \int \dk [\alpha(k)\alpha(-k) +2
  \beta(k) \alpha(-k)] \frac{G(k)}{1+fG(k)}~.
\end{equation}
Finally, using the correspondence formula \eqref{corr}, we obtain the
one- and two-point function of the perturbed boundary field theory,
\begin{align}
\label{ex1pt}
  \langle \Op(k) \rangle &= - \beta(k) \frac{G(k)}{1+fG(k)}~,\\
\label{ex2pt}
  \langle \Op(k) \Op(-k) \rangle -\langle \Op(k) \rangle \langle
  \Op(-k) \rangle&= \frac{G(k)}{1+fG(k)}~.
\end{align}
Comparing eqn.\ \eqref{ex1pt} with eqn.\ \eqref{exbc} justifies the
identification of $\cp$ with $\Op$, 
and both, eqns.\ \eqref{ex1pt} and \eqref{ex2pt}, are in agreement
with a field theory analysis. In particular, eqn.\ \eqref{ex2pt}
correctly reproduces Dyson's formula. 

In conclusion, we have presented an improved AdS/CFT correspondence
formula, which is able to account for multi-trace perturbations of the
boundary field theory. The calculation of the bulk on-shell action, in
particular holographic renormalization, is carried out as usual. 
The change occurs in the relation of the prescribed boundary data of
the bulk fields to the source of the boundary field theory.
We expect that most existing AdS/CFT
calculations in the literature are unaffected by the new formula,
since they involve single-trace perturbations, but new results for
multi-trace perturbations are anticipated. The improved method
might also be useful to construct holographic correspondence formulae for
non-asymptotically AdS spaces.

\section*{Acknowledgements}
I would like to thank E.~Witten for helpful comments. Financial
support by by the European Commission RTN programme
HPRN-CT-2000-00131, in which I am  associated with INFN, Sezione di
Frascati, is gratefully acknowledged.

\end{document}